\documentclass[a4paper]{article}

\usepackage[english]{babel}
\usepackage[utf8x]{inputenc}
\usepackage[T1]{fontenc}

\usepackage[a4paper,top=3cm,bottom=2cm,left=3cm,right=3cm,marginparwidth=1.75cm]{geometry}

\usepackage{amsmath}
\usepackage{graphicx}
\usepackage[colorinlistoftodos]{todonotes}
\usepackage[colorlinks=true, allcolors=blue]{hyperref}
\title{Future strategies for the discovery\\ and the precise measurement\\ of the Higgs self coupling
}
\author{Alain Blondel, Patrick Janot}
\date{}

\begin{document}
\maketitle

\begin{abstract}
The European Strategy for Particle Physics (ESSP) submitted in 2013 a deliberation document~\cite{deliberationDocument} to the CERN council explaining that a lepton collider with {\it "energies of 500\,GeV or higher could explore the Higgs properties further, for example the {\rm [Yukawa]} coupling to the top quark, the  {\rm [trilinear]} self-coupling  and  the  total  width."}. In view of the forthcoming ESPP update in 2020, variations on this qualitative theme have been applied, inaccurately, to the case of the ILC ~\cite{LCBReport,ICFAStatement}, to argue that an upgrade to 500\,GeV would allow the measurement of the Higgs potential and would increase the potential for new particle searches. As a consequence, the strategic question was raised again whether the FCC-ee design study ought to consider a 500\,GeV energy upgrade. In this note, we revisit the ESSP 2013 statement quantitatively and find 
\begin{itemize}
\item that the FCC-ee can measure the total width of the Higgs boson with a precision of 1.3\% -- the best precision on the market -- with runs at $\sqrt{s} = 240$, $350$, and $365$\,GeV, and without the need of an energy upgrade to 500\,GeV;

\item that the top Yukawa coupling will have been determined at HL-LHC at the $\pm$2.5\% level, albeit with some model dependence, without the need of 500\,GeV ${\rm e^+e^-}$ collisions; and that the combination of this HL-LHC result with the FCC-ee absolute Higgs coupling and width measurements breaks the model dependence, without the need of an energy upgrade to 500\,GeV; 

\item that, with the run plan presented for the ILC, the trilinear Higgs self-coupling can be inferred with a $3\sigma$ significance from the double-Higgs production cross-section measurement at the ILC500 after three decades of operation; but that the FCC-ee provides a similar sensitivity in 15 years, from the precise measurement of the single-Higgs production cross section as a function of $\sqrt{s}$, without the need of an energy upgrade to 500\,GeV; 

\item that the same FCC-ee, with four experiments instead of two, might well achieve the first model-independent $5\sigma$ demonstration of the existence of the trilinear Higgs self-coupling, while a centre-of-mass energy of about 1\,TeV or more is required for a linear collider to reach a similar sensitivity in a reasonable amount of time; and that a precise measurement of the trilinear Higgs self-coupling at the few per-cent precision level can realistically only be provided by the combination of FCC-ee and FCC-hh, being beyond reach of lepton colliders with a centre-of-mass energy up to at least 3\,TeV. 

\end{itemize}

On the new particle search front, the run plan of the FCC-ee includes an electroweak precision measurement program that is sensitive to new physics scales up to 70\,TeV. The Z factory run of the FCC-ee with $5\times 10^{12}$ Z decays can discover particles (e.g., dark matter or heavy neutrinos) that couple with a strength down to as little as $10^{-11}$ of the weak coupling. Finally, the FCC-hh potential for new particle searches at high mass exceeds that of any proposed linear collider project.

We conclude that 500 GeV is not a particularly useful energy for the lepton colliders under consideration, especially for the FCC-ee. A $5\sigma$ demonstration of the existence of the Higgs self-coupling is within reach at the energies foreseen for the FCC-ee, with a moderate change of configuration, which certainly deserves consideration.

\end{abstract}
\vfill\eject

\section{Operation models}

\begin{figure}[h]
\centering
\includegraphics[width=0.42\textwidth]{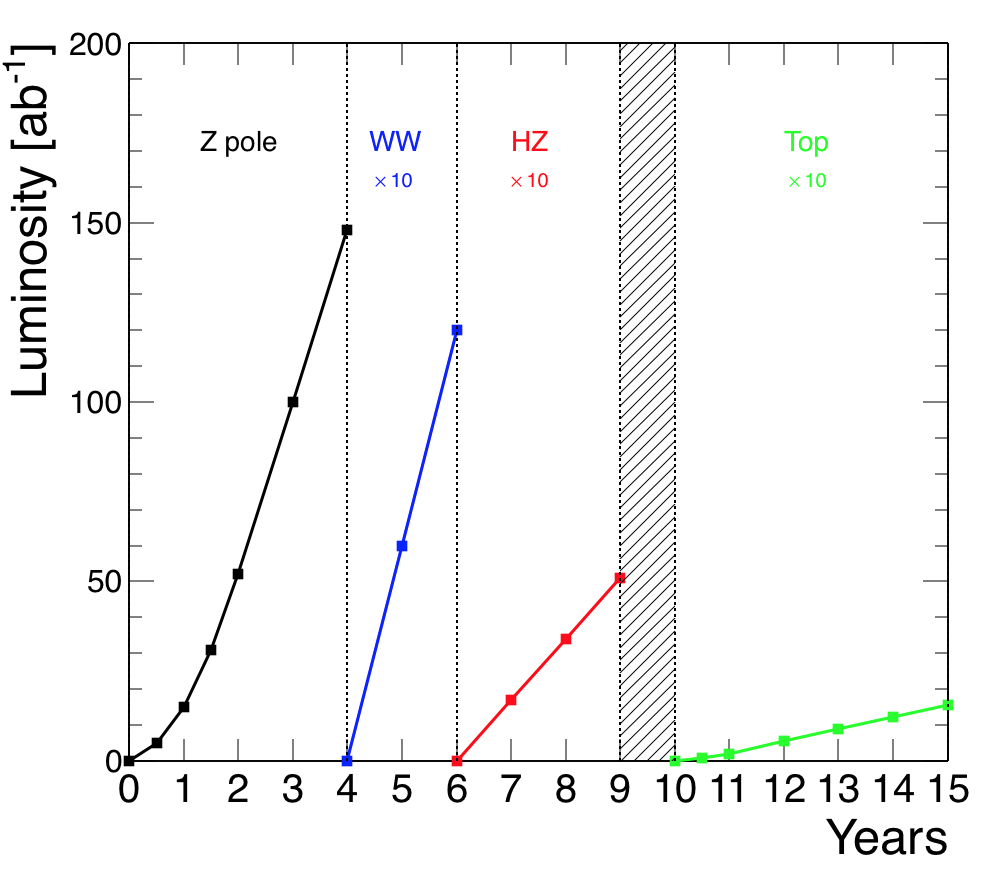}
\includegraphics[width=0.50\textwidth]{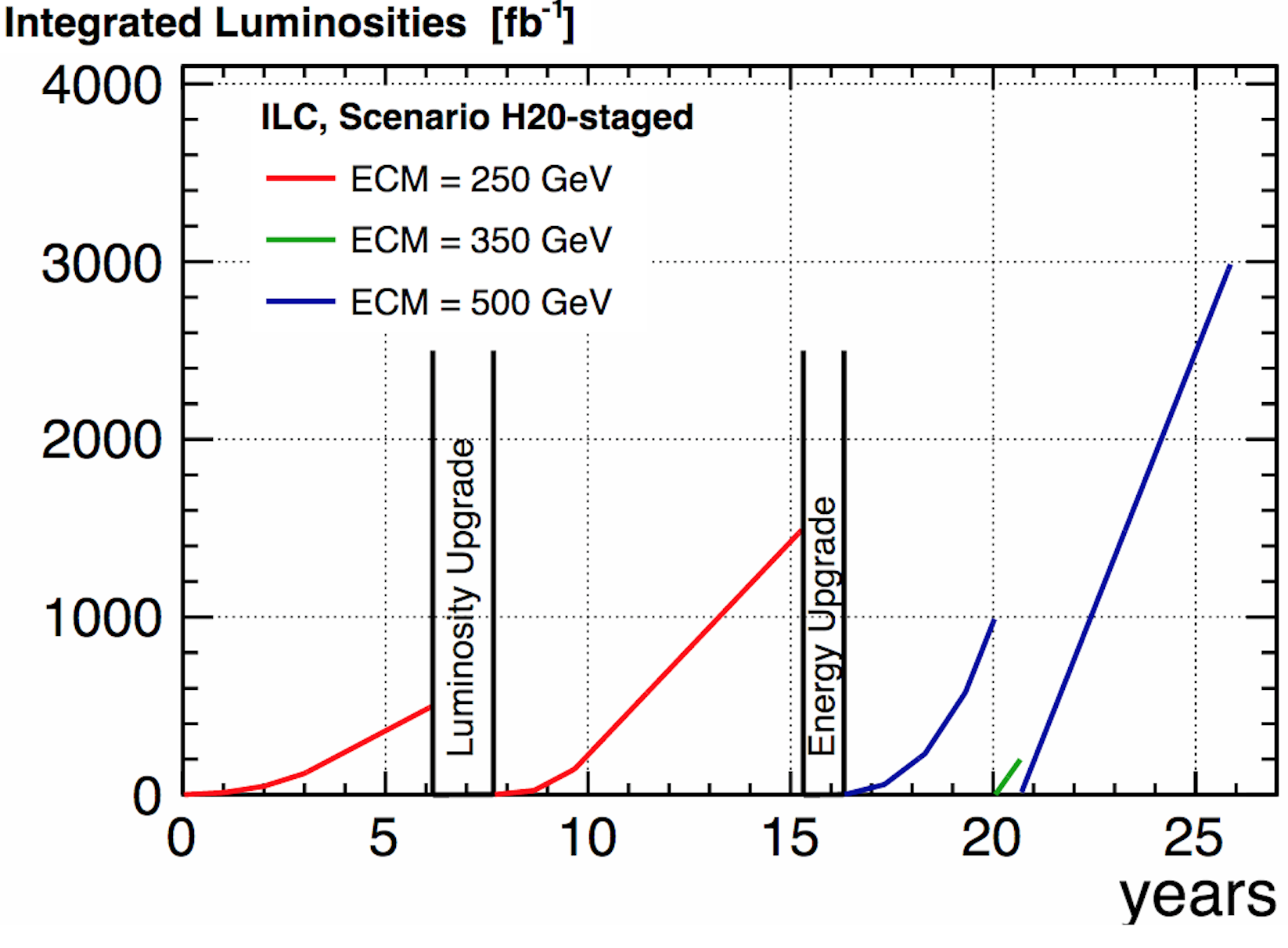}
\caption{\label{fig:OperationModel} \small Operation models for the FCC-ee (left) and the ILC (right). Note that both horizontal and vertical scales are different. The definitions of a typical year are also different: it conservatively corresponds to $1.2\times 10^7$\,seconds at the FCC-ee, while a more aggressive year of  $1.67\times 10^7$\,seconds is chosen for the ILC. The same definition would result either in a reduction of the FCC-ee operation to 10 years, or in an increase of the ILC operations to 40 years. The switches from one detector to another (push-pull) is not included in the duration of ILC operation.}
\end{figure}

The operation models of the FCC-ee~\cite{Benedikt:2651299} and the ILC~\cite{Fujii:2017vwa} are displayed in Fig.~\ref{fig:OperationModel}. 

In the first 15 years of operation, and with $\sim 1.67\times 10^7$\,seconds per typical year of running, the ILC will be able to accumulate $2\,{\rm ab}^{-1}$ at a single centre-of-mass energy, $\sqrt{s} = 250$\,GeV and with a single interaction point. In the most optimistic projection, the ILC might start data taking in 2032, so that only the first 10\% of the luminosity in that sample would be recorded in parallel with the HL-LHC data taking. 

\textbf{In its 15 years of operation, and with a more conservative running year of ${\bf 1.19\times 10^7}$ seconds, the FCC-ee will have measured the properties of the Z, W, Higgs and top with unprecedented precision}, with $150\,{\rm ab}^{-1}$ at the Z pole, $12\,{\rm ab}^{-1}$ at the WW threshold, $5\,{\rm ab}^{-1}$ at $\sqrt{s} = 240$\,GeV, $0.2\,{\rm ab}^{-1}$ at the top-pair threshold, and $1.5\,{\rm ab}^{-1}$ at $\sqrt{s} = 365$\,GeV, collected at two interaction points.

After this 15-years first stage at the EW scale, both projects propose access to the high-energy frontier: the FCC-ee tunnel is designed to subsequently host the FCC-hh, a hadron collider with $\sqrt{s} = 100$\,TeV; and the ILC infrastructure can be extended to deliver ${\rm e^+e^-}$ collisions at 500\,GeV. 

\section{Performance as Higgs factories}

In its baseline run plan, the FCC-ee operates as a Higgs factory for seven years, of which three at 240\,GeV and four at 365\,GeV. The expected statistical uncertainties on the various cross section times branching fractions measured at these energies~\cite{Benedikt:2651299,Gomez-Ceballos:2013zzn} are listed in Table~\ref{tab:HiggsMeasurements}. From these measurements, the Higgs couplings and total decay width can be fit 
in the so-called $\kappa$ framework~\cite{LHCHiggsCrossSectionWorkingGroup:2012nn,Heinemeyer:2013tqa} or in the truly model-independent manner in the SMEFT dimension-six operator framework~\cite{Grzadkowski:2010es}. The results of the $\kappa$ fit are summarized in Table~\ref{tab:FitResults} and are compared to the same fit applied to the HL-LHC projections~\cite{Cepeda:2650162} and to those of the ILC~\cite{Barklow:2017suo} after fifteen years at 250\,GeV. The SMEFT fit gives similar results~\cite{Barklow:2017suo}. 

The FCC-ee accuracies are subdivided in three categories: the first sub-column give the results of the fit expected with $5\,{\rm ab}^{-1}$ at 240\,GeV,  
the second sub-column -- directly comparable to the ILC column -- includes the additional $1.5\,{\rm ab}^{-1}$ at $\sqrt{s} = 365$\,GeV; and the last sub-column shows the result of the combined fit with HL-LHC projections. The fit to the stand-alone HL-LHC projections (first column) requires two additional assumptions to be made: here, the branching ratios into ${\rm c\bar c}$ and into exotic particles are set to their SM values. Instead, the fit to the HL-LHC projections becomes assumption-free when constrained with the FCC-ee measurements. 

\begin{table}[ht!]
\begin{center}
\caption{\small Relative statistical uncertainty on $\sigma_{\rm HZ} \times {\rm BR}({\rm H \to XX})$ and $\sigma_{\rm \nu\bar\nu H} \times {\rm BR}({\rm H \to XX})$, as expected from the FCC-ee data. All numbers (in \%) indicate 68\% C.L. intervals, except for the 95\% C.L. sensitivity in the last line. The accuracies expected with $5\,{\rm ab}^{-1}$ at 240\,GeV are given in the middle column, and those expected with $1.5\,{\rm ab}^{-1}$ at $\sqrt{s} = 365$\,GeV are displayed in the last column.\label{tab:HiggsMeasurements}}
\begin{tabular}{|l|r r|r r|}
\hline 
$\sqrt{s}$ (GeV) & \multicolumn{2}{|c|}{$240$} & \multicolumn{2}{|c|}{$365$} \\ \hline
Luminosity (${\rm ab}^{-1}$) & \multicolumn{2}{|c|}{5} & \multicolumn{2}{|c|}{$1.5$} \\ \hline
$\delta (\sigma {\rm BR}) / \sigma {\rm BR}$ (\%) & HZ & $\nu\bar\nu$ H &  HZ & $\nu \bar \nu$ H \\  \hline
${\rm H \to any}$          & $\pm 0.5$ &           & $\pm 0.9$    &             \\ 
${\rm H \to b \bar b}$     & $\pm 0.3$ & $\pm 3.1$ & $\pm 0.5$    & $\pm 0.9$   \\ 
${\rm H \to c \bar c}$     & $\pm 2.2$ &           & $\pm 6.5$    & $\pm 10$    \\ 
${\rm H \to gg}$           & $\pm 1.9$ &           & $\pm 3.5$    & $\pm 4.5$   \\ 
${\rm H \to W^+W^-}$       & $\pm 1.2$ &           & $\pm 2.6$    & $\pm 3.0$   \\ 
${\rm H \to ZZ}$           & $\pm 4.4$ &           & $\pm 12$     & $\pm 10$    \\ 
${\rm H \to \tau\tau} $    & $\pm 0.9$ &           & $\pm 1.8$    & $\pm 8$     \\ 
${\rm H \to \gamma\gamma}$ & $\pm 9.0$ &           & $\pm 18$     & $\pm 22$    \\ 
${\rm H \to \mu^+\mu^-}$   & $\pm 19$  &           & $\pm 40$     &             \\ 
${\rm H \to invis.}$       & $<0.3$    &           & $<0.6$       &             \\ \hline
\end{tabular} 
\end{center}
\end{table}

\begin{table}[h]
\begin{center}
\caption{\label{tab:FitResults} \small Precision determined in the $\kappa$ framework on the Higgs boson couplings and total decay width, as expected from the FCC-ee running as a Higgs factory for seven years, compared to those from HL-LHC and from the ILC running as a Higgs factory for 15 years.   All numbers (in \%) indicate 68\% C.L. sensitivities, except for the last line which gives the 95\% C.L. sensitivity on the "exotic" branching fraction, accounting for final states that cannot be tagged as SM decays. }
\begin{tabular}{|l|r|r|r|r|r|}
\hline Collider & {\small HL-LHC} & {\small ILC$_{250}$} & \multicolumn{3}{|c|}{FCC-ee$_{240+365}$} \\ \hline
Lumi (${\rm ab}^{-1}$) & {\small 3} & {\small 2} & $5_{240}$ & $\oplus 1.5_{365}$ & {\small $\oplus$ HL-LHC} \\ \hline
Years & {\small 10} & {\small 15} & 3 & $+$4 & \\ \hline
$\delta\Gamma_{\rm H}/\Gamma_{\rm H}$ (\%) & {\small 50} & {\small 3.6} & 2.7 & {\bf 1.3} & {\bf 1.1} \\ \hline
$\delta g_{\rm HZZ}/g_{\rm HZZ}$ (\%) & {\small 1.5} & {\small 0.3}  & 0.2 & {\bf 0.17}  & {\bf 0.16} \\ 
$\delta g_{\rm HWW}/g_{\rm HWW}$ (\%) & {\small 1.7} & {\small 1.7} & 1.3& {\bf 0.43}  &  {\bf 0.40} \\ 
$\delta g_{\rm Hbb}/g_{\rm Hbb}$ (\%) & {\small 3.7} & {\small 1.7} & 1.3 & {\bf 0.61}  &  {\bf 0.56} \\ 
$\delta g_{\rm Hcc}/g_{\rm Hcc}$ (\%) & {\small \bf SM} & {\small 2.3} & 1.7 & {\bf 1.21}  &  {\bf 1.18} \\ 
$\delta g_{\rm Hgg}/g_{\rm Hgg}$ (\%) & {\small 2.5} & {\small 2.2} & 1.6& {\bf 1.01}  &  {\bf 0.90} \\ 
$\delta g_{\rm H\tau\tau}/g_{\rm H\tau\tau}$ (\%) & {\small 1.9} & {\small 1.9} & 1.4 & {\bf 0.74}  &  {\bf 0.67}\\ 
$\delta g_{\rm H\mu\mu}/g_{\rm H\mu\mu}$ (\%) & {\small 4.3} & {\small 14.1} & 10.1 & {\bf 9.0}  & {\bf 3.8} \\ 
$\delta g_{\rm H\gamma\gamma}/g_{\rm H\gamma\gamma}$ (\%) & {\small 1.8} & {\small 6.4} & 4.8 & {\bf 3.9}  & {\bf 1.3} \\ 
$\delta g_{\rm Htt}/g_{\rm Htt}$ (\%) & {\small 3.4} & -- & -- & -- &  {\bf 3.1} \\ \hline
BR$_{\rm EXO}$ (\%) & ${\small \bf SM (0.0)}$ & {$\small < 1.7$} & $ < 1.2 $ & ${\bf < 1.0}$  & ${\bf < 1.0}$ \\ \hline
\end{tabular} 
\end{center}
\end{table}

\parskip=0.15cm
In addition to the unique electroweak precision measurement programme
at the Z pole and the WW threshold, sensitive to new physics scales up to 70\,TeV~\cite{Benedikt:2651299}, \textbf{the FCC-ee also provides the best model-independent precisions for all couplings accessible from Higgs boson decays, among all the  ${\bf e^+e^-}$ collider projects at the EW scale.} With larger luminosities delivered to several detectors at different centre-of-mass energies, \textbf{the FCC-ee improves over the model-dependent HL-LHC precisions by large factors for all non-rare decays, and breaks the model dependence inherent to hadron colliders.} Quantum corrections to Higgs couplings are at the level of a few \% in the standard model. \textbf{With a sub-per-cent precision for all these decays,  the FCC-ee is therefore able to test the quantum nature of the Higgs boson, without the need of a costly ${\bf e^+e^-}$ centre-of-mass energy upgrade. The FCC-ee also determines the Higgs boson width with a precision of 1.3\%}, which in turn allows the HL-LHC measurements to be interpreted in a model-independent way with the best accuracies.

\section{The value of a 500\,GeV upgrade}
\parskip=0.15cm

The precise measurement of the top-quark properties does not require ${\rm e^+e^-}$ or hadron colliders to be operated at the energy frontier. With a luminosity of $0.2\,{\rm ab}^{-1}$ recorded in a scan of the ${\rm t \bar t}$ threshold (below 350\,GeV),  the top-quark mass and width is best determined at the FCC-ee with statistical precisions  of $\pm 17$\,MeV and $\pm 45$\,MeV, respectively. The large systematic uncertainty from the knowledge of the strong coupling constant, which affects the measurement at linear colliders, is made negligible at the FCC-ee by the direct $\alpha_{\rm S}$ determination at the Z pole and the WW threshold~\cite{Gomez-Ceballos:2013zzn}. The largest FCC-ee centre-of-mass energy, $\sqrt{s} = 365$\,GeV, is chosen to optimally measure the top-quark electroweak couplings, without the need of incoming beam polarization~\cite{Janot:2015yza}. 

On the other hand, several Higgs boson couplings are not directly accessible from its decays, either because the masses involved, and therefore the decay branching ratios, are too small to allow for an observation within $10^6$ events -- as is the case for the couplings to the particles of the first SM family: electron, up quark, down quark -- or because the masses involved are too large for the decay to be kinematically open -- as is the case for the top-quark Yukawa coupling and for the trilinear Higgs boson self coupling. 

Traditionally, bounds on the top Yukawa and Higgs cubic couplings are extracted from the (inclusive and/or differential) measurements of the ${\rm t\bar t H}$ and ${\rm HH}$ production cross sections, which require significantly higher centre-of-mass energies, either in ${\rm e^+e^-}$ or in proton-proton collisions. The measurement of these two couplings is often used as an argument to demonstrate the absolute necessity of ${\rm e^+e^-}$ collisions at $\sqrt{s}=500$\,GeV. The validity of the argument is examined next.

\subsection{The Top Yukawa coupling}

With enough luminosity at 500\,GeV accumulated over a second period of 12 years, the top Yukawa coupling can indeed be determined with a precision of $\pm 6.3\%$ at the ILC500~\cite{Fujii:2015jha}, improved to $\pm 4\%$ with $\sqrt{s}=550$\,GeV, where the ${\rm t\bar t H}$ production cross section is largest. The ${\rm t\bar t H}$ production, however, has already been detected at the LHC with a significance larger than $5\sigma$ by both the ATLAS~\cite{Aaboud:2018urx} and CMS~\cite{Sirunyan:2018hoz} collaborations, which constitutes the first observation of the top-quark Yukawa coupling. \textbf{It is therefore expected that the top Yukawa coupling will be determined with a similar or better precision already by the HL-LHC (conservatively ${\bf \pm 3.4\%}$, with some model dependence~\cite{Cepeda:2650162}), 30 years before the ILC500. This precision will be further constrained to ${\bf \pm 3.1}\%$ through the model-independent combined fit with FCC-ee data (Table~\ref{tab:FitResults}).} 

The FCC-ee also has access to this coupling on its own, through its effect at quantum level on the ${\rm t\bar t}$ cross section just above production threshold, $\sqrt{s} = 350$\,GeV.  The FCC-ee measurements at lower energies are important to fix the value of the strong coupling constant $\alpha_{\rm S}$. This precise measurement  allows the QCD effects to be disentangled from those of the top Yukawa coupling at the ${\rm t\bar t}$ vertex. A precision of $\pm 10\%$ is achievable at the FCC-ee on the top Yukawa coupling. The comparison of this measurement and that from ${\rm t\bar t H}$ production at the HL-LHC is a valuable test of quantum theory with loops involving Higgs bosons. 

Furthermore, \textbf{the FCC-hh has the potential to reach a precision better than ${\bf \pm 1\%}$ with the measurement of the ratio of the ${\bf t\bar t H}$ to the ${\bf t\bar t Z}$ cross sections~\cite{fcc-hh-projections}, when combined with the top EW couplings precisely measured at the FCC-ee~\cite{Janot:2015mqv}}.  

\subsection{The trilinear Higgs self coupling}

The determination of the Higgs self-interactions is of primary importance. They characterize the Higgs potential, whose structure is intimately connected to the naturalness problem and to the question of the (meta)stability of the EW vacuum. Moreover, they control the properties of the electroweak phase transition, determining its possible relevance for baryogenesis. Sizeable deviations in the Higgs self-couplings are expected in several BSM scenarios, including for instance Higgs portal models or theories with Higgs compositeness. They however remain intangible at the LHC: at present, the trilinear Higgs coupling is loosely constrained at the ${\cal O}(10)$ level, and the high-luminosity LHC program will only test it with an ${\cal O}(1)$ accuracy~\cite{Cepeda:2650162}. The prospects for extracting the quadrilinear Higgs self-coupling are even less promising. 

\textbf{At the energy frontier, only the FCC-hh has the potential to reach a precision of the order of ${\bf \pm 5\%}$ in the determination of the trilinear Higgs coupling from double-Higgs production, in combination with the precise Higgs decay branching ratio measurements from the FCC-ee.} The highest-energy ${\rm e^+e^-}$ colliders (beyond 1~TeV) are limited to a precision of about $\pm 10$-$15\%$~\cite{Abramowicz:2016zbo, clic-projections}, and so is the high-energy upgrade of the LHC \cite{fcc-hh-projections}. At lower energies and after 27 years of operation  (15 years at 250~GeV and 12 years at 500~GeV), the ILC would reach a $3 \sigma$ sensitivity to the trilinear Higgs coupling~\cite{Fujii:2015jha}, {\it i.e.} a precision of about $\pm 30\%$.

\begin{figure}[htbp]
\begin{center}
\includegraphics*[width=0.50\textwidth]{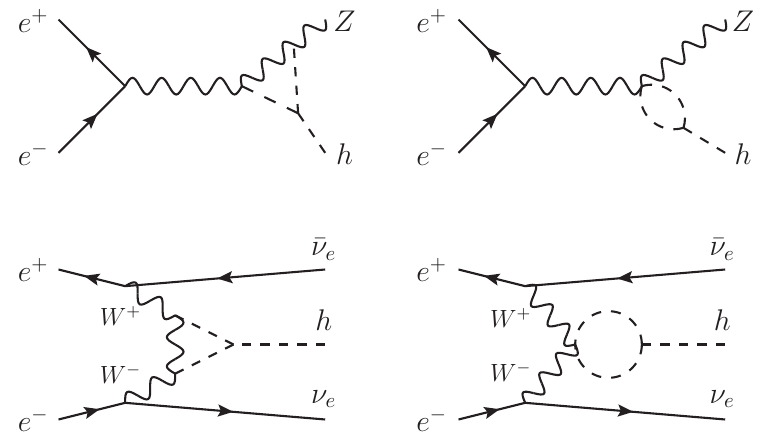}
\includegraphics*[width=0.36\textwidth]{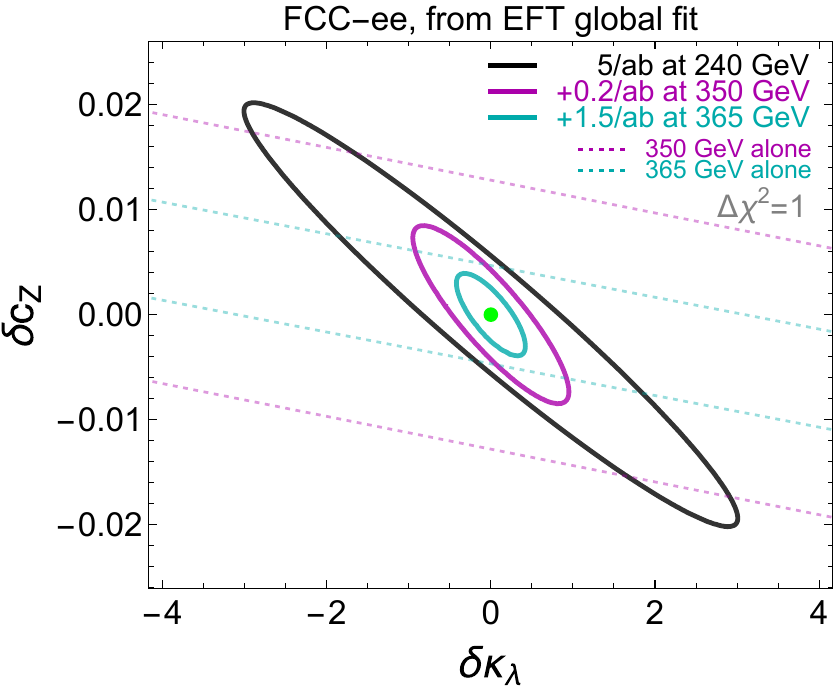}
\end{center}
\vskip -0.25cm \caption{\label{fig:h3-TGC-atEE} \small Left, from Ref.~\cite{DiVita:2017vrr}: sample Feynman diagrams illustrating the effects of the Higgs trilinear self-coupling on single Higgs process at next-to-leading order. Right: standalone FCC-ee precision in the simultaneous determination of the Higgs trilinear self-coupling $\kappa_\lambda$ and the HZZ/HWW coupling $c_{\rm Z}$, at 240\,GeV (black ellipse), 350\,GeV (purpled dashed), 365\,GeV (green dashed), and by combining data at 240 and 350\,GeV (purple ellipse), and at 240, 350, and 365\,GeV (green ellipse).}
\end{figure}

On the other hand, with the large luminosity delivered at 240 and 365\,GeV, the FCC-ee proposes an early and privileged sensitivity to the trilinear coupling. This sensitivity is provided by its centre-of-mass-energy-dependent effects at the quantum level on single Higgs observables~\cite{McCullough:2013rea}, such as the HZ and the $\nu\bar\nu$H  production cross sections, representative diagrams of which are displayed in Fig.~\ref{fig:h3-TGC-atEE} (left). For example, a 100\% modification of the Higgs self-coupling changes the HZ cross section by 2\% at 240\,GeV and 0.5\% at 365\,GeV~\cite{DiVita:2017vrr}. Robust and model-independent bounds can be obtained~\cite{DiVita:2017vrr, Maltoni:2018ttu, deBlas} through a global (Higgs and EW) dimension-six operator fit, that includes in particular the Higgs self-coupling $\kappa_\lambda$ and the coupling to SM gauge bosons $c_{\rm Z}$. \textbf{When all the FCC-ee centre-of-mass energies are included in the fit, a model-independent precision of ${\bf \pm 42\%}$ can be achieved on  ${\bf \kappa_\lambda}$ (Fig.~\ref{fig:h3-TGC-atEE}, right), reduced to ${\bf \pm 34\%}$ in combination with HL-LHC, and to ${\bf \pm 12\%}$ when only $\mathbf{\kappa_\lambda}$ is allowed to vary.} The FCC-ee EW measurements at lower energies are equally important to fix extra parameters that would otherwise enter the Higgs fit and open flat directions that cannot be resolved~\cite{deBlas}. 

\vskip -0.1cm

It is interesting to note that the doubling of integrated luminosity expected if the ring were arranged to accommodate four detectors instead of two would be instrumental to this measurement. With four IPs, two years at the Z pole and one year at the WW threshold would suffice to get the same integrated luminosity as after six years with two IPs. The saved years and the four detectors could then be optimally used to accumulate about 12\,${\rm ab}^{-1}$ at 240\,GeV and 5.5\,${\rm ab}^{-1}$ at 350 and 365\,GeV. {\bf The larger data sample would yield a model-independent measurement of $\kappa_\lambda$ with a precision of ${\bf \pm 25\%}$, reduced to ${\bf \pm 21\%}$ in combination with the HL-LHC, and to ${\bf \pm 9\%}$ if only $\kappa_\lambda$ is allowed to vary. The first ${\bf 5\sigma}$ demonstration of the existence, i.e. the discovery, of the Higgs self-coupling is within reach in 15 years at the FCC-ee.}

The measurements of fundamental SM parameters via their quantum effects are often called "indirect", as opposed to "direct" measurements from on-shell production, with the implicit innuendo (which is sometimes incorrect) that direct measurements would be more model independent than their indirect counterparts. The double Higgs direct production, however, includes Feynman graphs that do not depend on $\kappa_\lambda$ and that interfere with graphs that do. The measurement of this cross section at high-energy colliders can thus only be interpreted as a determination of the trilinear Higgs coupling in an indirect manner. The ultra-precise measurements of the HZ and $\nu\bar \nu$H cross sections performed at the FCC-ee are no different in this respect, and give information similar to double Higgs production with 500\,GeV ${\rm e^+e^-}$ collisions.  The measurement of the double Higgs production cross section at the FCC-hh with an unequalled precision is of course essential to the accurate characterisation of the Higgs potential. 
\section{Conclusion}

The next accelerator project must allow the broadest possible field of research. This is the case for the FCC. To begin with, the FCC-ee was successfully demonstrated after five years of conceptual design study to achieve performance superior to that assumed for the original "TLEP"~\cite{Gomez-Ceballos:2013zzn}, albeit with two interaction points instead of four. The FCC-ee would measure the Z, W, Higgs, and top properties in ${\rm e^+e^-}$ collisions, either for the first time or with orders of magnitude  improved precision, thereby giving access to either much higher scales or much smaller couplings. The FCC-ee is the most powerful of all proposed ${\rm e^+e^-}$ colliders at the EW scale --- all things being equal, in particular the duration of operations (Fig.~\ref{fig:OperationModel}). The FCC-ee proposes a broad, multifaceted exploration to: 
\begin{enumerate}
\item Measure a comprehensive set of electroweak and Higgs observables with high precision;
\item Tightly constrain a large number of the parameters of the standard model;
\item Unveil small but significant deviations with respect to the standard model predictions;
\item Observe rare new processes or particles, beyond the standard model expectations;
\end{enumerate}
and therefore maximizes opportunities for major fundamental discoveries. The combination of the FCC-ee with the HL-LHC is superior to the ILC500 (or to its combination with the HL-LHC), in particular for the measurement of the top Yukawa coupling. With 15 years of FCC-ee operations between 90 and 365\,GeV, this combination also provides a $3\sigma$ sensitivity to the trilinear Higgs self coupling, similar to that obtained after three decades of ILC operation at 250 and 500\,GeV. 

The FCC-ee also meets the last part of the ESPP 2013 guideline about ${\rm e^+e^-}$ colliders\footnote{"There  is  a  strong  scientific  case  for  an  electron-positron  collider,  complementary  to  the  LHC, that  can  study  the  properties  of  the  Higgs  boson  and other particles with unprecedented precision and  whose  energy  can  be  upgraded."~\cite{ESPP2013}}  in the most ambitious manner, as the FCC-ee tunnel is designed to subsequently host the FCC-hh, a hadron collider with a centre-of-mass energy of 100\,TeV. Combined with the FCC-ee measurements, the FCC-hh physics reach at the energy and precision frontiers exceeds that of any proposed linear collider energy upgrade. 

A precise measurement at the few per-cent level of the trilinear Higgs self-coupling would require to run a (multi-)TeV lepton collider either for a prohibitive duration (to integrate a luminosity typically a factor five larger than currently planned) or with a centre-of-mass energy well above 3\,TeV. Today, the only realistic possibility of an accurate measurement is offered by the combination of the FCC-ee and FCC-hh data. The FCC might also be the only hope for a first $3\sigma$ observation of the quartic Higgs self-coupling~\cite{Fuks:2017zkg}.

An upgrade of the FCC-ee to 500\,GeV would be costly in time and funds for the FCC program in general, while its physics output would be trivially superseded by the FCC-hh. For a linear collider to reach a $5\sigma$ sensitivity on the Higgs self-coupling requires an even more costly upgrade to almost 1\,TeV. A much more efficient upgrade would be to increase the number of FCC-ee experiments from two to four, which would be the fastest and most cost-effective way -- by doubling the delivered luminosity and by using at 240 and 365\,GeV the running time saved at the Z pole and the WW threshold -- to achieve the first $5\sigma$ demonstration of the existence of the trilinear Higgs self-coupling. Should this goal be identified as a priority by the forthcoming update of the European Strategy for Particle Physics, a study of an FCC-ee with four interaction points could be envisioned in the context of the FCC technical design study. 

\section*{\small Acknowledgements} 
\vskip -0.25cm \small We are indebted to Jorge de Blas, Christophe Grojean, Jiayin Gu, and Matthew McCullough for enlightening discussions, expert suggestions, subtle comments, and running fits throughout the development of this analysis. We sincerely thank Frank Zimmermann for starting an inspiring discussion about the need of a 500\,GeV upgrade for the FCC-ee, and therefore motivating us to write this note. Credit is also due to Michael Benedikt, Philippe Bloch, and Daniel Treille for a careful reading of the manuscript and for their wise comments. Hats off to Michael Peskin for stressing that the FCC-ee precision EW measurements should be (and indeed are) essential for a precise and model-independent determination of the Higgs self-coupling. 

\vfill\eject
\bibliographystyle{jhep}
\bibliography{sample}

\end{document}